\documentclass[twocolumn,floats,floatfix,nofootinbib,superscriptaddress]{revtex4-1}
\usepackage[applemac]{inputenc}
\usepackage{amsmath,amssymb}
\usepackage{graphicx}
\usepackage{float}
\usepackage{hyperref}

\begin{document}
%\title{Spherical collapse of scalar fields: a Galerkin-Collocation multidomain approach}

\title{Distorted black holes: a characteristic view}

%\author{M. A. Alcoforado}
%\email{malcoforado@hotmail.com}
%\affiliation{Departamento de F\'{\i}sica Te\'orica - Instituto de F\'{\i}sica
	%A. D. Tavares, Universidade do Estado do Rio de Janeiro, 
	%R. S\~ao Francisco Xavier, 524. Rio de Janeiro, RJ, 20550-013, Brazil}

%\author{W. O. Barreto}
%\email{wobarreto@gmail.com}
%\affiliation{Departamento de F\'{\i}sica Te\'orica - Instituto de F\'{\i}sica
%	A. D. Tavares, Universidade do Estado do Rio de Janeiro, 
%	R. S\~ao Francisco Xavier, 524. Rio de Janeiro, RJ, 20550-013, Brazil}
%\affiliation{Centro de F\'{\i}sica Fundamental, Universidad de Los Andes, M\'erida 5101,  Venezuela}

\author{H. P. de Oliveira}
\email{henrique.oliveira@uerj.br}
\affiliation{Departamento de F\'{\i}sica Te\'orica - Instituto de F\'{\i}sica A. D. Tavares, Universidade do Estado do Rio de Janeiro, R. S\~ao Francisco Xavier, 524. Rio de Janeiro, RJ, 20550-013, Brazil}

\date{\today}

\begin{abstract}
%We present a simple domain decomposition code based on the Galerkin-Collocation method to integrate the field equations of the Bondi problem. The algorithm is stable, exhibits exponential convergence when considering the Bondi formula as an error measure, and is computationally economical. We have incorporated features of both Galerkin and Collocation methods along with the establishment of two non-overlapping subdomains. We have further applied the code to show the decay of the Bondi mass in the nonlinear regime and its power-law late time decay. Another application is the determination of the wave-forms at the future null infinity connected with distinct initial data.
We investigate the interaction between a non-rotating black hole and incoming gravitational waves using the characteristic formulation of the Einstein field equations, framed as a Bondi problem. By adopting retarded time as the null coordinate and recognizing that the final state is invariably a black hole, we demonstrate that an apparent horizon forms once sufficient mass accretes onto the black hole. We derive the evolution of the Bondi mass and compute its final value, enabling us to quantify the fraction of the incident mass absorbed by the black hole. Additionally, we establish a scaling law for the absorbed mass as a function of initial parameters, such as the amplitude of the gravitational wave data. Furthermore, we explore the dynamics when a reflecting barrier surrounds the black hole. For low-amplitude initial waves, the barrier reflects the waves, leaving the original black hole as the end state. Conversely, high-amplitude waves lead to the formation of an apparent horizon that engulfs the barrier, producing a new, larger black hole.
\end{abstract}

\maketitle

%%%%%%%%%%%%%%%%%%%%%%
\section{Introduction}
%%%%%%%%%%%%%%%%%%%%%%

The interaction of gravitational waves with a black hole, which results in a distorted black hole, emerged as an important problem when numerical relativity began to effectively address general relativistic issues beyond spherical symmetry. One of the main motivations for studying this problem is that highly distorted black holes may represent the late stages of the merger between two black holes. Additionally, other aspects of distorted black holes that are of interest include the characteristics of the emitted waveforms, the dynamics of the apparent horizons, and the efficiency of gravitational wave extraction. Furthermore, distorted black holes have provided a valuable theoretical framework for developing efficient computational codes to solve the nonlinear field equations first in axisymmetric spacetimes and subsequently in more general spacetimes.

The investigation of distorted black holes started with constructing initial configurations representing black holes interacting with Brill waves around them. This work included both nonrotating black holes \cite{abrahams_et_al_92,aninos_et_al_94,brandt_seidel_95_II,camarda_seidel_98,camarda_seidel_99,baker_et_al_2000,brown_lowe_2004,hpoliveira_rodrigues} and rotating black holes \cite{brandt_seidel_95_II,brandt_seidel_96,brandt_et_al_2003}, examined in axisymmetric and more general three-dimensional setups \cite{camarda_seidel_98,camarda_seidel_99,brandt_et_al_2003}. The dynamics of these distorted black holes were addressed in references \cite{aninos_et_al_94,brandt_seidel_95_I,brandt_seidel_95_II,brandt_seidel_96,camarda_seidel_98,camarda_seidel_99,baker_et_al_2000}, which focused on the waveforms emitted during the process.  Another important aspect discussed was the dynamics of apparent horizons that form around strongly perturbed black holes. In all these studies, the numerical codes implemented relied on the Cauchy formulation of the field equations.

An alternative approach is to study the dynamics of distorted black holes through the characteristic initial value problem \cite{bondi,sachs,winicour_lrr}. The groundbreaking work of Bondi et al. \cite{bondi} laid the foundation for the characteristic formulation when an isolated source emits gravitational waves. Sachs \cite{sachs} later extended this formulation for general three-dimensional spacetimes. Many authors have utilized the characteristic formulation, employing both outgoing and ingoing light cone foliations, to investigate nonlinear spherical and nonspherical perturbations of black holes \cite{gomez_et_al_97,gomez_et_al_94,crespo_deoliveira,papadopoulos,nerozzi,barreto}. Specifically, references \cite{papadopoulos, nerozzi, barreto} focus on distorted black holes using the characteristic formulation with ingoing light cones.

In the present paper, we examined distorted black holes using a characteristic formulation with outgoing light cone foliations that differ from previous works on this topic. This approach allowed us to track the decay of the Bondi mass in accordance with the Bondi formula until the formation of an apparent horizon, enabling us to determine the amount of mass absorbed by the black hole. We derived a scaling relation between the absorbed mass and the initial strength of the gravitational wave, demonstrating that this scaling law is independent of the specific initial data. Additionally, we investigated how modifying the inner boundary by introducing a reflecting barrier  \cite{gomez_et_al_94,crespo_deoliveira} around the black hole affects its dynamics.

The paper is organized as follows: In Section 1, we present the fundamental equations and specify the boundary conditions, adopting the Bondi frame \cite{bondi} to describe the asymptotic behavior of the relevant functions. Section 2 briefly outlines the numerical method used to solve the field equations. We present our numerical results in Section 3, highlighting the decay of the Bondi mass consistent with the Bondi formula and the observed scaling behavior of the absorbed mass. In the final section, we discuss new directions for future research on this topic.

%%%%%%%%%%%%%%%%%%%%%%%%%%%%%%
\section{The field equations}%
%%%%%%%%%%%%%%%%%%%%%%%%%%%%%%

%The metric established by Bondi, van der Burgh and Metzner \cite{bondi} that describes axisymmetric and asymptotic spacetimes takes the form

The metric established by Bondi, van der Burgh, and Metzner \cite{bondi}, which describes axisymmetric and asymptotic spacetimes, is expressed as follows

\begin{eqnarray}
ds^2=-\left(\frac{V}{r}\mathrm{e}^{2\beta}-U^2 r^2 \mathrm{e}^{2\gamma}\right) du^2 - 2\mathrm{e}^{2\beta} du dr \nonumber \\
\nonumber \\
- 2 U r^2 \mathrm{e}^{2\gamma} du d\theta + r^2(\mathrm{e}^{2\gamma} d \theta^2 + \mathrm{e}^{-2\gamma}\sin^2 \theta d\varphi^2). \label{eq3}
\end{eqnarray}

\noindent Here, $u$ is the retarded time coordinate such that $u=constant$ denotes the outgoing null cones; the radial coordinate $r$ is chosen demanding that the surfaces $(u,r)$ have area equal to $4 \pi r^2$ and the angular coordinates $(\theta,\varphi)$ are constant along the outgoing null geodesics. The metric functions $\gamma, \beta, U$ and $V$ depend on the coordinates $u,r,\theta$ and satisfy the vacuum field equations $R_{\mu\nu} = 0$. 

The simplest non-trivial solution of the field equations is the Schwarzschild solution given by

\begin{eqnarray}
	\beta&=&U=\gamma=0, \nonumber \\
	\\
	V&=&r-2M_0 \nonumber,
\end{eqnarray}

\noindent where $M_0$ is the black hole mass and $r \geq 2 M_0$.

%Our objective is to study the nonlinear interaction of gravitational waves surrounding a Schwarzschild black hole with mass $M_0$. To this aim, it is necessary to solve the field equations numerically in the characteristic formalism, but before presenting schematically the overall procedure, some appropriate definitions are required. We express the function $V(u,r,\theta)$ as

Our goal is to study the nonlinear interaction of gravitational waves surrounding a Schwarzschild black hole with mass $M_0$. To this end, we will numerically solve the field equations within the framework of the characteristic formalism. Before presenting the procedure schematically, some essential definitions are required. We express the function $V(u,r,\theta)$ as

\begin{equation}
	V(u,r,\theta) = r - 2M_0+S(u,r,\theta).
\end{equation}

\noindent Since $2 M_0$ is a natural scale of the spacetime, it motivates the introduction of a new radial coordinate $\eta$ by

\begin{equation}
	r=2M_0(1 + \eta),
\end{equation}

\noindent where the original $r \geq 2M_0$ now corresponds to $\eta \geq 0$, effectively excising the black hole.  Substituting the new radial coordinate into Eq. (3) and dividing by $2 M_0$, we obtain

\begin{equation}
	\bar{V}(u,\eta,\theta) = \eta +\bar{S}(u,\eta,\theta),
\end{equation}

\noindent where $\bar{V}=V/2M_0$ and $\bar{S}=S/2M_0$. We then define $\bar{U}=2M_0 U$ and introduce an auxiliary variable $\bar{Q}$: 

\begin{equation}
	\bar{Q} \equiv (1+\eta)^4 \mathrm{e}^{2(\gamma-\beta)}\bar{U}_{,\eta}.
\end{equation}

%\noindent Instead of the function $\gamma(u,r,\theta)$ we introduce $\Gamma(u,r,\theta)$ given by

\noindent  Rather than using the function $\gamma(u,\eta,\theta)$, we introduce a new variable $\Gamma(u,\eta,\theta)$, given by

\begin{equation}
	\Gamma(u,\eta,\theta) \equiv (1+\eta) \gamma(u,\eta,\theta).
\end{equation}

%We can express the field equations in dimensionless variables with the above definitions. In the framework of the characteristic formulation, the field equations are divided into two sets, the hypersurface, and the evolution equations, given in a compact form as 

With these definitions, the field equations can be expressed in dimensionless variables. In the characteristic formulation \cite{winicour_lrr}, the field equations are divided into two sets: the hypersurface equations and the evolution equations, which are compactly written as:

\begin{eqnarray}
	\beta_{,\eta} &=&
	\mathcal{H}_\beta(\Gamma), \\
	 %\frac{1}{2}(1+r)\left(\frac{\Gamma}{1+r}\right)_{,r}^2 = \mathcal{H}_\beta(\Gamma), \\
	\nonumber \\
	%\left[(1+r)^2 Q\right]_{,r}&=&\mathcal{H}_Q(\beta,\Gamma),  \\
	Q_{,\eta}&=&\mathcal{H}_Q(\beta,\Gamma),  \\
	\nonumber \\
	U_{,\eta}&=&\mathcal{H}_U(\beta,\Gamma,Q), \\
	\nonumber \\
	S_{,\eta}&=&\mathcal{H}_S(\beta,\Gamma,U), \\
	\nonumber \\
	\Gamma_{,\tilde{u} \eta}&=&\mathcal{H}_\Gamma(\beta,\Gamma,U,S),
\end{eqnarray}

%\noindent For the sake of convenience we removed the bar of the metric functions. The $\mathcal{H}$'s are functions of the Bondi variables dictated by the field equations that we show in the Appendix. 

\noindent where the retarded time is rescaled as $\tilde{u}=u/2M_0$. For convenience, we have removed the bars from the metric functions. The terms $\mathcal{H}$ represent functions of the Bondi variables, as dictated by the field equations, which are provided in detail in the Appendix A.

%In writing the above equations, we have considered the normalized variables $\beta/\sin^4 \theta \rightarrow \beta$, $\Gamma/\sin^2 \theta \rightarrow \Gamma$, $U/\sin \theta \rightarrow U$ and $Q/\sin \theta \rightarrow Q$, recalling that all variables are dimensionless. In the last equation, the retarded time is rescaled as $u/2M_0 \rightarrow u$.

%The spatial domain comprises $0 \leq r < \infty$, where $r=0$ denotes the original black hole horizon and $0 \leq \theta \leq \pi$. The boundary conditions imposed on the field variables are distinct from those imposed in the collapse of pure gravitational waves (see Refs. \cite{varios}). In summary, we have that all fields vanish at $r=0$ according to

The regularity conditions at the symmetry axis require that $\gamma /\sin^2 \theta$ and $U/\sin \theta$ be continuous at $\theta = 0, \pi$. Consequently, the same conditions apply to $\Gamma$ and $Q$. In the numerical code, we adopt normalized variables: $\beta/\sin^4 \theta \rightarrow \beta$, $\Gamma/\sin^2 \theta \rightarrow \Gamma$, $U/\sin \theta \rightarrow U$, and $Q/\sin \theta \rightarrow Q$. %Additionally, the retarded time is rescaled as $u/ 2M_0 \rightarrow u$.

The boundary conditions imposed on the field variables differ from those used when the spacetime does not initially contain a black hole, such as in the collapse of gravitational waves \cite{gomez_et_al_jmp_94} or a scalar field \cite{gomez_winicour_92}. In our case, all fields vanish at the black hole horizon $\eta=0$, as described by:

\begin{eqnarray}
	\Gamma &&= \mathcal{O}(\eta),\quad \beta = \mathcal{O}(\eta),\quad U = \mathcal{O}(\eta^2) \\
	\nonumber \\
	Q &&= \mathcal{O}(\eta),\quad S=\mathcal{O}(\eta^2). 
\end{eqnarray} 

\noindent The asymptotic conditions must ensure the asymptotic flatness of the spacetime. To achieve this, we adopt the Bondi frame \cite{bondi}, which gives

%We adopt the Bondi frame \cite{bondi} to dictate the asymptotic conditions that guarantee the asymptotic flat nature of the spacetime:

\begin{eqnarray}
	\Gamma &&= \Gamma_0+\mathcal{O}(\eta^{-1}), \\
	\nonumber \\
	\beta &&= -\frac{\Gamma_0^2}{4\eta^2}+\mathcal{O}(\eta^{-3}), \\
	\nonumber \\
	U &&= -\frac{(\Gamma_0 \sin^4 \theta)_{,\theta}}{\sin^3 \theta\,\eta^2} + \mathcal{O}(\eta^{-3}), \\
	%\nonumber \\
	%Q &&= \mathcal{O}(\eta), \\
	\nonumber \\
	S &&= -2\mathcal{M} + \mathcal{O}(\eta^{-1}),
\end{eqnarray} 

\noindent where $\Gamma_0=\Gamma_0(\tilde{u},\theta)$ and $\mathcal{M}=\mathcal{M}(\tilde{u},\theta)$ is the rescaled Bondi mass aspect. From the above relations, we obtain that $Q = \mathcal{O}(\eta)$.

%We evaluate the Bondi mass from the mass aspect that must include the initial mass of the black hole, $M_0$. Following Bondi et al. \cite{bondi},  the Bondi mass is %and recasting the original function $V(u,r,\theta)$ given by Eq. (3), the Bondi mass is 

We evaluate the Bondi mass using the mass aspect, which must account for the initial mass of the black hole, $M_0$. Following Bondi et al. \cite{bondi}, the Bondi mass is given by

\begin{equation}
	M_B(\tilde{u}) = M_0\left(1+\int_0^\pi \mathcal{M}(\tilde{u},\theta) \sin \theta \, d \theta\right).
\end{equation}

%\noindent In the absence of the interaction provided by the gravitational waves, $\mathcal{M} = 0$ and the Bondi mass is the black hole mass $M_0$. The Bondi mass satisfies the well-known Bondi formula, connecting its variation with the action of the news function identified as $\mathcal{N}(u,\theta)=\partial \Gamma_{0}/\partial u$. Then, from Ref. \cite{bondi}, we have

\noindent In the absence of interaction with gravitational waves, $\mathcal{M}=0$, and the Bondi mass reduces to the black hole mass $M_0$. The Bondi mass satisfies the well-known Bondi formula, which relates its variation to the action of the news function, identified as 
\begin{equation}
	\mathcal{N}(\tilde{u},\theta)=\frac{\partial \Gamma_{0}}{\partial \tilde{u}}.
\end{equation}

\noindent. From Ref. \cite{bondi}, we obtain

\begin{equation}
	\frac{d M_B}{d\tilde{u}} = -M_0\,\int_0^\pi 
	\mathcal{N}^2(\tilde{u},\theta) \sin \theta \, d \theta.
\end{equation}

%\noindent The verification of the Bondi formula constitutes a valid numerical test to certify the convergence of the numerical code we describe in the sequence. 

\noindent Verifying the Bondi formula provides a reliable numerical test to confirm the convergence of the numerical code that we describe in the sequence.

%%%%%%%%%%%%%%%%%%%%%%%%%%%%%%%
\section{The numerical method}%
%%%%%%%%%%%%%%%%%%%%%%%%%%%%%%%

We briefly describe the spectral code based on the Galerkin-Collocation method \cite{alcoforado_et_al_2022} to integrate the field equations (8) - (12). First, it is convenient to introduce computational angular and radial coordinates, $x,y$ by

\begin{eqnarray}
	\theta = \arccos x,\\
	\nonumber \\
	\eta = L_0 \left(\frac{1+y}{1-y}\right),
\end{eqnarray}

\noindent such that $-1 \leq x \leq 1$ and $-1 \leq y \leq 1$ correspond to  $0 \leq \theta \leq \pi$ and  $0 \leq \eta < \infty$, respectively; $L_0$ is the map parameter.

To construct the radial basis functions for the spectral approximations, we use the rational Chebyshev polynomials \cite{boyd} defined by
%The radial basis functions are, in general, linear combinations of the rational Chebyshev polynomials \cite{boyd}

\begin{equation}
	TL_k(\eta) \equiv T_k\left(y=\frac{\eta-L_0}{\eta+L_0}\right),
\end{equation}

\noindent where $ T_k (y) $ is the Chebyshev polynomial of $ k$-th order. We combine the rational Chebyshev polynomials, allowing the basis functions to satisfy the boundary conditions (13) - (18). The Legendre polynomials $P_j(x)$ are the angular basis functions with distinct parity about $x=0$ ($\theta=\pi/2$) depending on the metric function. Starting with $\Gamma$ as an even function of $x$, $\beta$ and $S$ will also be even functions $x$, while $U$ and $Q$ will be odd functions of $x$.

For the sake of illustration, we present the spectral approximation of the function $\beta$:

\begin{equation}
	\beta(\tilde{u},\eta,x) = \sum_{k}^{N_\eta^{(\beta)}}\,\sum_{j=0}^{N_x^{(\beta)}}\,\hat{\beta}_{kj}(\tilde{u}) \Psi_k^{(\beta)}(\eta) P_{2j}(x),
\end{equation}

\noindent where $\hat{\beta}_{kj}$ are the unknow coefficients or modes, $N_\eta^{(\beta)}, N_x^{(\beta)}$ are the radial and angular truncation orders. We have implemented the radial basis function, $\Psi_k^{(\beta)}(\eta)$, such that asymptotically decay as $\mathcal{O}(\eta^{-2})$ satisfying the condition (16). Interestingly, we noticed that $\Psi_k^{(\beta)}(\eta) \sim \mathcal{O}(1)$ close to the original black hole horizon is numerically more efficient than if we have set the condition (13), that is instead of $\mathcal{O}(\eta)$.

The spectral approximation of the function $U$ has the same basis function of $\beta$, but the angular basis functions are the odd Legendre polynomial, $P_{2 j+1}(x)$. The expansion of $Q$ has $\eta TL_k(\eta)$ as the radial basis functions and $P_{2 j+1}(x)$ as the angular basis functions. Considering the remaining functions, $S$ and $\Gamma$, in both spectral approximations, we have the same basis functions $\Psi_k(\eta) P_{2j}(x)$, where $\Psi_k(\eta) \sim \mathcal{O}(\eta)$ near $\eta=0$ and $\Psi_k(\eta) \sim \mathcal{O}(1)$ asymptotically. 

In each spectral approximation, we established the corresponding modes (see Appendix B), namely $\hat{U}_{kj}(\tilde{u}),\,\hat{Q}_{kj}(\tilde{u}),\,\hat{S}_{kj}(\tilde{u}),\,\hat{\Gamma}_{kj}(\tilde{u})$. The following steps to determine these modes are: $(i)$ we obtain the residual equations after substituting the spectral approximations into the field equations; and $(ii)$ we impose that the residual equations vanish at the collocation points. By applying the same strategy of Ref. \cite{alcoforado_et_al_2022}, we employ the G-NI method to vanish the residual equation related to the evolution equation in an average sense \cite{finlayson,boyd,canuto}. As a consequence, the field equations (8) - (11) become algebraic equations to evaluate the modes  $\hat{\beta}_{kj}(\tilde{u}),\,\hat{Q}_{kj}(\tilde{u}),\,\hat{U}_{kj}(\tilde{u}),\,\hat{S}_{kj}(\tilde{u})$, respectively. Eq. (12) results in a set of ordinary differential equations, or a dynamical system,  for the modes $\hat{\Gamma}_{kj}(\tilde{u})$.

Taking advantage of the hierarchical structure provided by the characteristic scheme, we engendered relations involving the truncation orders that reflect this hierarchy. Starting with $N_\eta$ and $N_x=N_\eta/2$, the radial and angular truncation orders present in the spectral approximation of $\Gamma$, we adopt for simplicity $N_\eta^{(\beta)}=N_\eta^{(Q)}=N_\eta^{(U)} = N_\eta+2$,  $N_\eta^{(S)}= N_\eta+2$ and similar relations for the angular truncation orders.

%\begin{eqnarray}
%N_r^{(\beta)}=N_r^{(Q)}=N_r^{(U)} = \frac{12}{10}N_r,\\
%\nonumber \\
%N_x^{(\beta)}=N_x^{(Q)}=N_x^{(U)} = \frac{12}{10}N_x, \\
%\nonumber \\
%N_r^{(S)}= \frac{13}{10}N_r, \quad N_x^{(S)} = \frac{13}{10}N_x.
%\end{eqnarray}   

We start the spacetime evolution with the initial configuration provided by

\begin{equation}
	\Gamma(\tilde{u}=\tilde{u}_0,\eta,x) = \epsilon \eta (1-x^2) \mathrm{e}^{-(\eta-\eta_0)^2/\sigma^2},
\end{equation}

\noindent where $\epsilon$ plays the role of the initial amplitude, $\eta_0$ is the center of the wave distribution and $\sigma$ its width. In all simulations, we set $\eta_0=3$ and $\sigma=1$. After translating the initial data (26) in terms of the spectral modes $\hat{\Gamma}_{kj}(\tilde{u}_0)$, we integrate the dynamical system with the Cash-Karp adaptative stepsize integrator \cite{CK}.

By choosing the initial data with an even function of $x$, the field equations demand that for any $\tilde{u}$, the metric functions $\Gamma, \beta$ and $S$ are even functions of $x$, while $U, Q$ are odd functions of $x$. Consequently, we have placed the angular collocation points at $0 \leq x \leq 1$ instead of the whole angular domain.

\begin{figure}[htb]
	\includegraphics[width=7.5cm,height=6.0cm]{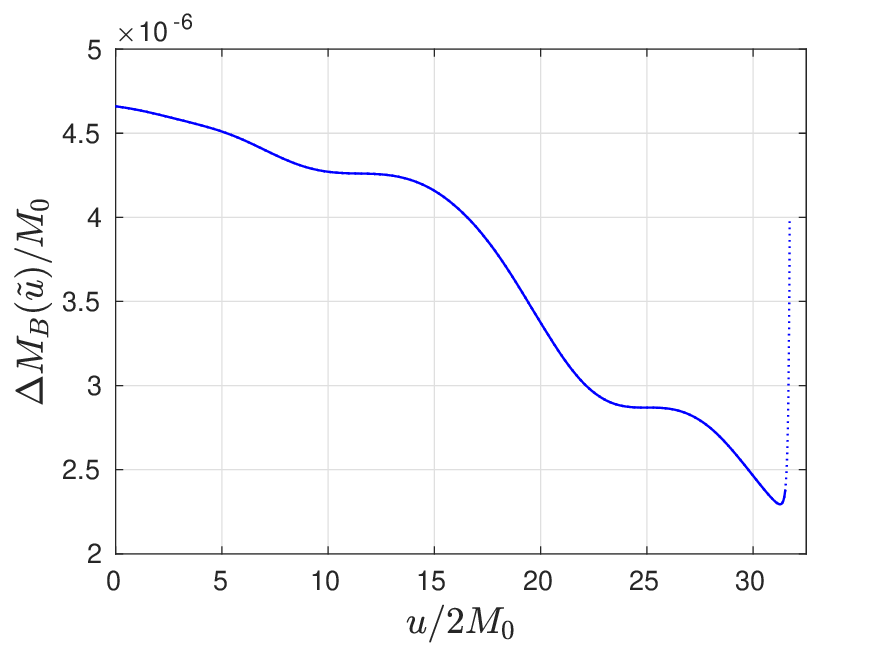}
	\includegraphics[width=7.5cm,height=6.0cm]{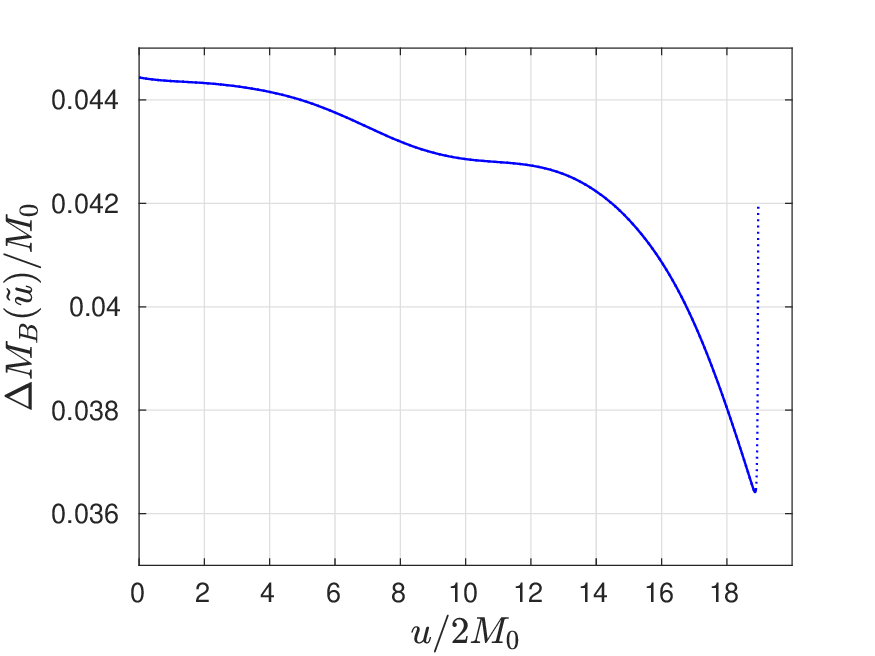}
	\caption{Decay of $\Delta M_B(\tilde{u})/M_0$ for $\epsilon=0.001$ (upper panel) and $\epsilon=0.1$ (lower panel), where the map parameter is $L_0=0.2$ and $\tilde{u}=u/2M_0$. The resolution is $N_\eta=120,\,N_x=60$.}
\end{figure}

%%%%%%%%%%%%%%%%%%%%%%%%%%%%
\section{Numerical Results}%
%%%%%%%%%%%%%%%%%%%%%%%%%%%%

%Before presenting the numerical results, discussing a crucial physical aspect of the system is necessary. Irrespective of the amount of energy present in the gravitational waves surrounding the black hole, the end state will necessarily be another black hole with mass $M_0+\delta M_{\mathrm{abs}}$, where $\delta M_{\mathrm{abs}}$ is the mass absorbed by the black hole. It means that an apparent horizon always forms, which is signalized when the expansion of radial null rays vanish, or $\Theta = \mathrm{e}^{-2 \beta}/r$ vanishes, in other words, when $\beta \rightarrow \infty$ at some point. Therefore, the integration diverges when the apparent horizon forms, trapping some of the initial gravitational wave mass, expanding the excised region, and releasing another mass fraction. 

\subsection{Behavior of the Bondi mass}

Before presenting the numerical results, it is important to discuss a crucial physical aspect of the system. Regardless of the energy in the gravitational waves surrounding the black hole, the final state will always be another black hole with mass $M_0 + \delta M_{\mathrm{abs}}$, where $\delta M_{\mathrm{abs}}$ is the mass absorbed by the black hole. This implies that an apparent horizon will always form, signaled when the expansion of radial null rays vanishes, or $\Theta = \mathrm{e}^{-2 \beta}/\eta$ vanishes, meaning $\beta \to \infty$ at some point. Consequently, the integration diverges upon the horizon’s formation, trapping part of the gravitational wave mass, expanding the excised region, and releasing the rest.

From Eq. (19), we define $\Delta M_B(\tilde{u})$ by 

\begin{equation}
	\Delta M_B(\tilde{u}) \equiv M_B(\tilde{u})-M_0,
\end{equation} 

%\noindent Like the Bondi mass, this quantity decays until the numerical integration crashes, delimitating the amount of mass that falls into the hole. We present in Fig. 2 the decay of $\Delta M_B(u)$ per unit of the unperturbed black hole for $\epsilon=0.001$ corresponding to a tiny amount of mass disturbing the black hole and $\epsilon=0.1$ for a typical nonlinear evolution. The initial Bondi mass in both cases are $M_B(0) \simeq 1.00000466M_0$ and $M_B(0) \simeq 1.0443 M_0$, respectively.   

\noindent that is the mass due to the gravitational atmosphere around the black hole evaluated at $\tilde{u}$. Similar to the Bondi mass, $\Delta M_B(\tilde{u})$ decays until the numerical integration fails, delimitating the amount of mass that falls into the black hole. Figure 1 illustrates the decay of $\Delta M_B(\tilde{u})$ for two cases: $\epsilon = 0.001$, corresponding to a small perturbation, and $\epsilon = 0.1$, representing a typical nonlinear evolution. The initial Bondi masses in both cases are $M_B(0) \simeq 1.00000466 M_0$ and $M_B(0) \simeq 1.0443 M_0$, respectively.

\begin{figure}[htb]
	\includegraphics[width=7.5cm,height=6.0cm]{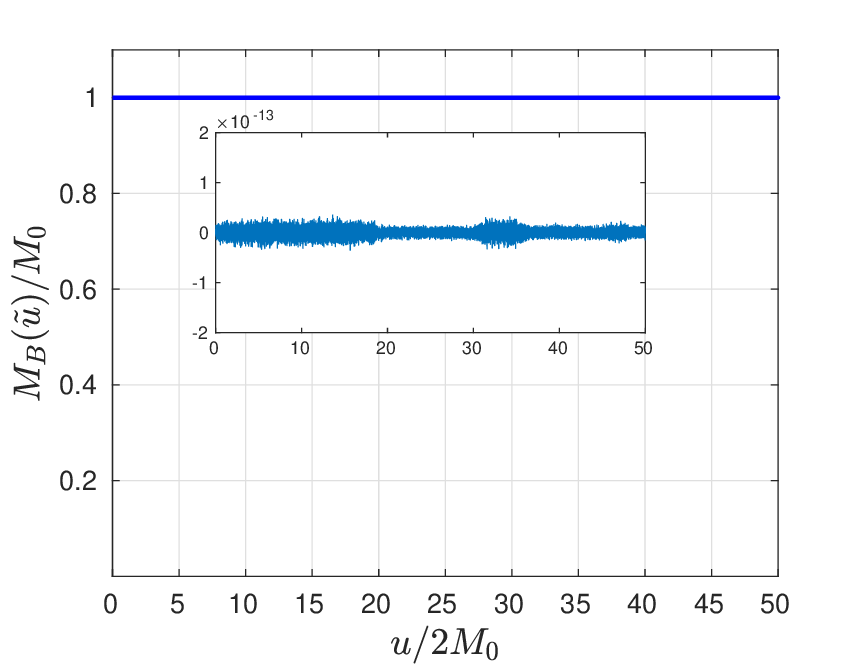}
	\caption{Evolution of the Bondi mass for $\epsilon=10^{-9}$ showing the it does not alter in time. Notice that the linear regime is recovered with $\Delta M_B \approx 0$ numerically.}
\end{figure}

%We point out that the exception is the case provided by the linear approximation where the perturbation due to the gravitational wave atmosphere produces a negligible amount of mass in the sense of $\Delta M_B(u) \approx 0$, which does not change the original black hole mass during the whole evolution. To test the linear approximation, we set $\epsilon = 10^{-9}$ yielding $\Delta  M_B(0)/M_0 \simeq \mathcal{O}(10^{-15})$ that is virtually the numerical zero. Fig. 3 shows the evolution of $M_B(u)/M_0$ and the inset $\Delta  M_B(0)/M_0$. In these instances, the code reproduced the expected behavior of the linear approximation. We have also reproduced the linear approximation for all initial amplitudes smaller than a minimum value, $\epsilon_{\mathcal{min}}$. For the initial data (26), $\epsilon_{\mathcal{min}} \approx 4.0 \times 10^{-7}$ rendering $\Delta  M_B(0)/M_0 \simeq \mathcal{O}(10^{-14})$.

One exception occurs in the linear regime, where the perturbation caused by the gravitational waves is negligible, leading to $\Delta M_B(\tilde{u}) \approx 0$, leaving the original black hole mass unchanged. For $\epsilon = 10^{-9}$, we find $\Delta M_B(0)/M_0 \simeq \mathcal{O}(10^{-15})$, which is virtually zero. Figure 2 shows the evolution of $M_B(\tilde{u})/M_0$ and, in the inset, $\Delta M_B(0)/M_0$. The code accurately reproduces the expected behavior of the linear approximation for all initial amplitudes below a minimum value, $\epsilon_{\mathrm{min}}$, where $\epsilon_{\mathrm{min}} \approx 4.0 \times 10^{-7}$ results in $\Delta M_B(0)/M_0 \simeq \mathcal{O}(10^{-14})$.

We briefly explore another exception to the above behavior when a reflecting barrier is placed around the black hole, as discussed in Refs. \cite{gomez_et_al_94,crespo_deoliveira}. The barrier reflects the gravitational waves as long as they are not strong enough to form an apparent horizon, leaving behind the black hole with the original mass $M_0$. When the horizon forms, the barrier falls inside the black hole.

The reflecting barrier is positioned at $r=R_0$ with $R_0 > 2M_0$. All fields vanish at the barrier except for the function $V(u,r,\theta)$, for which

\begin{equation} 
	V(u,r=R_0,\theta)=R_0-2M_0.
\end{equation}  

\noindent After introducing a new radial coordinate, $r=R_0(1+\eta)$, and the reescaled retarded time $\tilde{u}=u/R_0$, we obtain modified expressions for the Bondi mass and formula
\begin{eqnarray}
	M_B(\tilde{u}) &=& M_0\left(1+\frac{1}{K_0}\int_0^\pi \mathcal{M}(\tilde{u},\theta) \sin \theta \, d \theta\right),	\\
	\nonumber \\
	\frac{d M_B}{d\tilde{u}} &=& -\frac{R_0}{2}\,\int_0^\pi 
	\mathcal{N}^2(\tilde{u},\theta) \sin \theta \, d \theta,
\end{eqnarray}

\noindent where $K_0=2M_0/R_0$. For $R_0=2M_0$, we recover Eqs. (19) and (21). Figure 3 shows the evolution of $\Delta M_B(\tilde{u})$ with the reflecting barrier placed at different positions, indicated by $K_0=1.0,\, 0.998,\, 0.95$, and $0.93$. An apparent horizon forms for $K_0 = 1.0$, corresponding to the lower plot in Fig. 2, and also when the reflector is placed very close to the horizon at $K_0 = 0.998$ (or $R_0 \approx 2.004 M_0$. In these cases, part of the gravitational radiation, including the reflector, becomes trapped within the horizon. However, for $K_0 = 0.95$ and $K_0=0.93$, $\Delta M_B(u)$ decays continuously, indicating that the reflector radiates away the gravitational waves, leaving the original black hole as the remnant.

\begin{figure}[htb]
	\includegraphics[width=7.5cm,height=6.0cm]{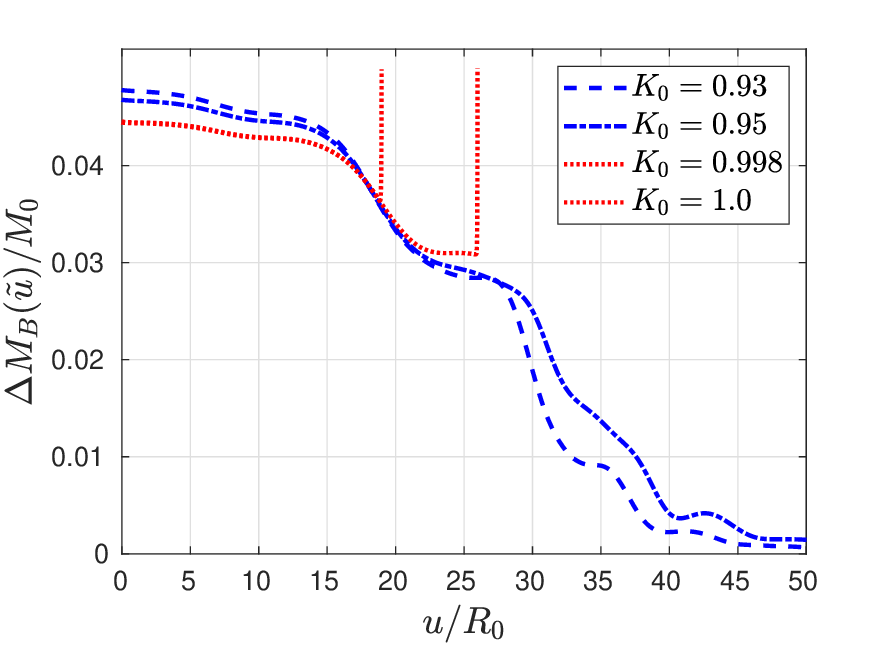}
	\caption{Decay of $\Delta M_B(\tilde{u})$ for $\epsilon=0.1$ with the reflecting barrier placed at distinct positions, $R_0=2.0 M_0,\,2.004 M_0,\,2.105 M_0$ and $2.15 M_0$ corresponding to $K_0=1.0,\, 0.998,\, 0.95$ and $0.93$, respectively. Note that $\tilde{u} = u/R_0$. }
\end{figure}

\begin{figure}[htb]
	\includegraphics[width=7.5cm,height=6.0cm]{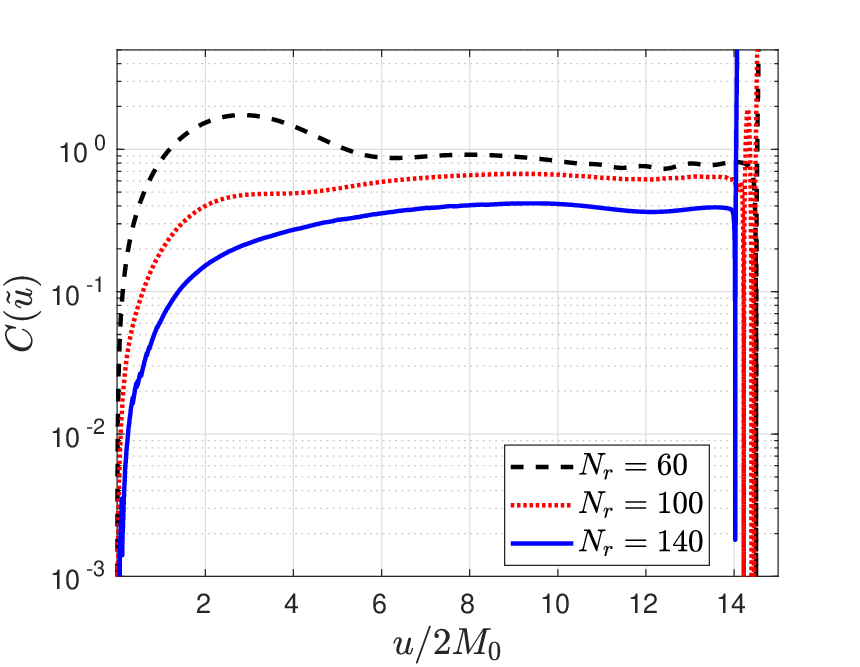}
    \includegraphics[width=7.5cm,height=6.0cm]{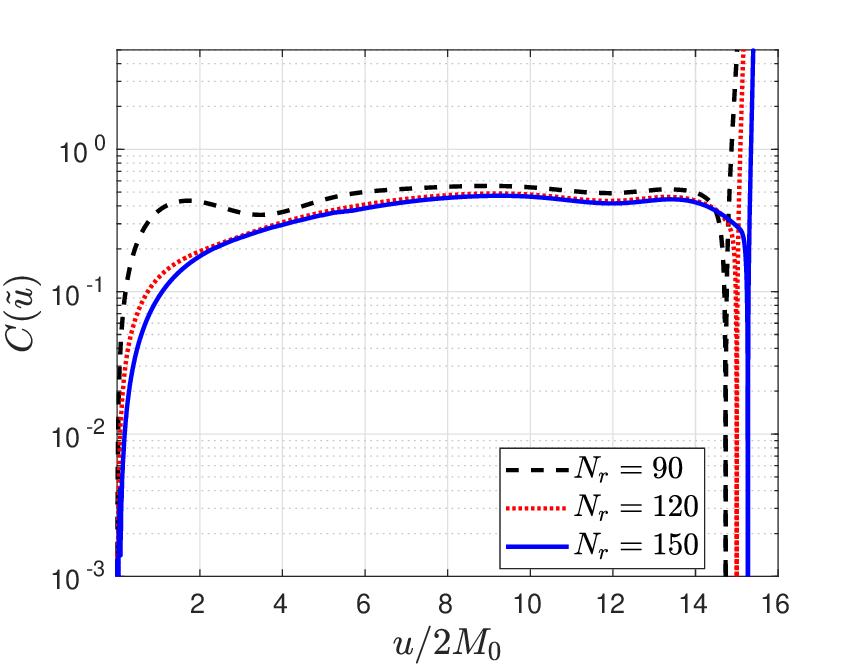}
	%\caption{Illustration of the convergence in the deviation of the global energy conservation. There is a clear decay of the error, where for the maximum resolution (continuous line) the error is less $0.03\%$ for most of the evolution.  }
	\caption{Qualitative illustration of the convergence in the deviation of the global energy conservation. Here, we have set the initial amplitude $\epsilon=0.2$ and the map parameter $L_0=0.2$. In the first panel, $N_x=N_\eta/2$, and in the second, $N_x=N_r\eta/3$. In both cases, the error in the global energy conservation decays with the increase of the resolution indicated by the continuous lines. We have obtained that $C_{\mathrm{max}} \simeq 0.3\%$.}
\end{figure}

\subsection{Convergence test}
We validate the code by verifying the Bondi formula expressing global energy conservation. %The procedure is to define the following function, $C(u)$ \cite{winicour}, that measures the relative error in the global energy conservation
The relative error in global energy conservation is measured by the following function, $C(\tilde{u})$ \cite{gomez_et_al_jmp_94} given by

\begin{eqnarray}
	C(\tilde{u}) &=& \bigg| \frac{M_B(0)-M_B(\tilde{u})}{M_B(0)} - \frac{1}{2 M_B(0)}\times \nonumber \\
	\nonumber \\
	&&\,\int_{-1}^{1}\,dx\,\int_0^{\tilde{u}}\,\mathcal{N}^2(\tilde{u},x)\,d\tilde{u} \bigg| \times 100,
\end{eqnarray}

\noindent where, for an exact evolution $C(\tilde{u})=0$. We calculate the spatial integral using a quadrature formula \cite{boyd} and the time integral with the trapezoidal rule with unequal steps. We present in Fig. 4 the behavior of $C(\tilde{u})$ for $\epsilon = 0.2$  with three distinct resolutions: $N_\eta=60, 100, 140$ with $N_x=N_\eta/2$, and $N_r=90, 120, 150$ with $N_x=N_\eta/3$. %In both cases, the map parameter is $L_0=0.2$. Since the numerical integration diverges at about $u/2M_0 \simeq 14.3$, it is inevitable the increase of $C(u)$; nevertheless, before the formation of the apparent horizon, the convergence with the increase of resolution is evident. 
In both cases, the map parameter is $L_0 = 0.2$. Since numerical integration diverges around $u/2M_0 \simeq 14.3$, an increase in $C(\tilde{u})$ is inevitable; however, before the apparent horizon forms, the convergence improves with increasing resolution.

\subsection{Scaling relation for the absorbed gravitational waves}

We are interested in studying the behavior of the absorbed mass, denoted by $\Delta M_B(\tilde{u}_f)$, as a function of the initial amplitude, $\epsilon$, where $\tilde{u}_f$ is the final retarded time immediately preceding the failure of integration. To facilitate the analysis, it is convenient to define the absorbed mass relative to the unperturbed black hole mass, $M_0$, as

\begin{equation}
	\delta M_{\mathrm{abs}} \equiv \frac{\Delta M_b(\tilde{u}_f)}{M_0} =  \frac{M_B(\tilde{u}_f)-M_0}{M_0}.
\end{equation}

\noindent Before presenting our numerical results, we first consider the case of a very small initial amplitude ($\epsilon \ll 1$). From the initial data, we have $\Gamma \sim \mathcal{O}(\epsilon)$, and the field equations dictate the following scaling relations

\begin{eqnarray}
	\beta \sim \mathcal{O}(\epsilon^4)\;\;U \sim \mathcal{O}(\epsilon^2),\;\;S \sim \mathcal{O}(\epsilon^2).
\end{eqnarray}

\noindent Since the mass aspect $\mathcal{M}(\tilde{u},\theta)$ is derived from the function $S$, we infer that $\Delta M_B(u) \sim \mathcal{O}(\epsilon^2)$. Consequently, the relative absorbed mass satisfies

\begin{eqnarray}
	\delta M_B(\tilde{u}_f) \sim \mathcal{O}(\epsilon^2).
\end{eqnarray}

\noindent This result suggests a power-law relationship between $\delta M_B(\tilde{u}_f)$ and the initial amplitude $\epsilon$, regardless of the specific choice of initial data provided $\epsilon \ll1 $.

We obtained the numerical data by computing the absorbed mass for various value of $\epsilon$. In addition to the initial data (26), we have considered initial data for a pure quadrupole mode 

\begin{equation}
\Gamma(\tilde{u}=0,\eta,x) = \epsilon\, \mathrm{e}^{-(\eta - \eta_0)^2/\sigma^2}, 
\end{equation} 

\noindent where we have set $\eta_0=1$ and $\sigma=1$, as well as

\begin{equation}
\Gamma(\tilde{u}=0,\eta,x) = \frac{\epsilon(1-x^2)}{(1+2\eta^2)}.
\end{equation} 

The numerical simulations were conducted using the resolution parameters: $N_\eta=180$, $N_x=60$, and map parameter $L_0=0.15$. Notably, as mentioned before, even for values of $\epsilon$ on the order of $\epsilon \sim \mathcal{O}(10^{-6})$, an apparent horizon forms, leading to a black hole with a small absorbed mass. Increasing the initial amplitudes revealed the following scaling law that best fits the numerical data for any initial data family

\begin{equation}
	\delta M_{\mathrm{abs}} = K_0\mathrm{\epsilon}^\nu,
\end{equation}

\noindent where $\nu \approx 2.052,\, 2.07$ and $2.19$, for the corresponding initial data sets (36), (26), and (35). This indicates a scaling relation that closely aligns with the theoretical prediction (34) for small initial amplitudes. Fig. 5 displays plots of the numerical data represented by symbols alongside the above scaling relation, illustrating the agreement between the theoretical and numerical distributions.

Additionally, it is also possible to fit the numerical data with a nonextensive relation \cite{tsallis,picolli} that generalizes the power-law (37): 

\begin{equation}
	\delta M_{\mathrm{abs}} = K_0\left[1-(1-\alpha \epsilon^\nu)^\mu\right],
\end{equation}

\noindent where again $K_0$ depends on the particular initial data set, and the remaining parameters are approximately equal with parameter values $\alpha \approx 0.1408,\,\nu \approx 2.0517$ and $\mu \approx 1.0134$. The parameter $K_0$, however, depends on the particular initial data family.

%The numerical simulations were carried out with resolution parameters $N_\eta=180$, $N_x=60$, and map parameter $L_0=0.15$. Notably, as mentioned before, even for values of $\epsilon$ on the order of $\epsilon \sim \mathcal{O}(10^{-6})$, an apparent horizon forms, leading to a black hole with a small absorbed mass. Extending the initial amplitudes to higher values revealed the following nonextensive law that best fits the numerical data for any initial data family

%\begin{equation}
%\delta M_{\mathrm{abs}} = K_0\left[1-(1-\alpha \epsilon^\nu)^\mu\right],
%\end{equation}

%\noindent with parameter values $\alpha \approx 0.1408,\,\nu \approx 2.0517$ and $\mu \approx 1.0134$. The parameter $K_0$, however, depends on the particular initial data family.

%\noindent that best fit the numerical for any initial data family data with the parameters values: $\alpha \approx 0.1408,\,\nu \approx 2.0517$ and $\mu \approx 1.0134$; only  $K_0$ depends on the particular initial data family.

\begin{figure}[htb]
	\includegraphics[width=7.5cm,height=6.0cm]{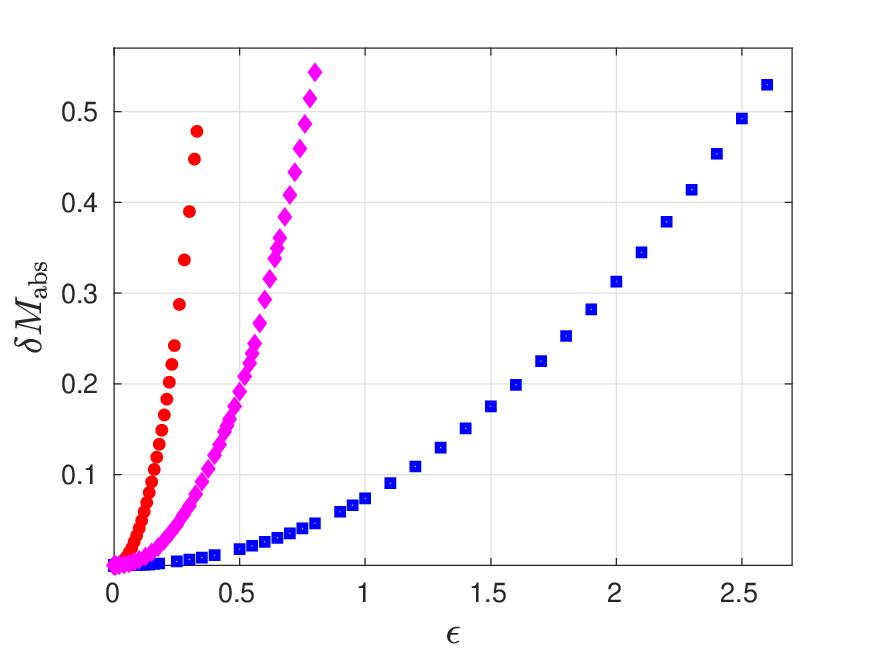}
	%\center{(a)}\\
	\includegraphics[width=7.5cm,height=6.0cm]{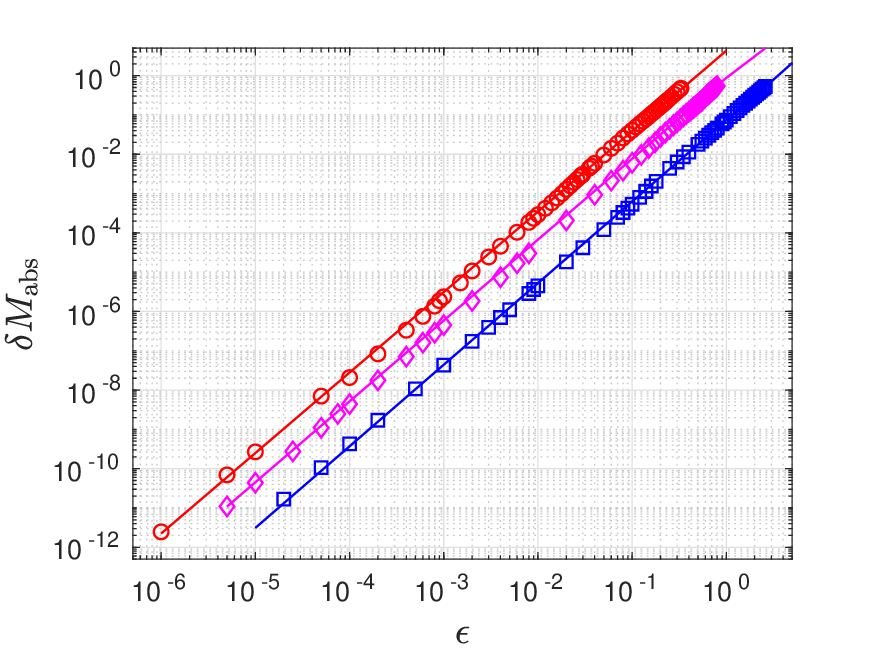}%{Mabs_loglog_plots.eps}
	%\center{(b)}
	\caption{Upper panel: absorbed mass $\delta M_{\mathrm{abs}}$ in function of the initial amplitude $\epsilon$ for the initial data (circles), (diamnonds), and (squares). Lower panel: log-log plots of the numerical distributions and the theoretical power-law scaling relation (37).}
\end{figure}

\section{Final considerations}

In this paper, we have explored the interaction of gravitational waves with a Schwarzschild black hole using the formulation of the classic Bondi problem. The field equations were solved numerically with the spectral Galerkin-Collocation method. In contrast with the previous works on this subject, we adopted the retarded time $u$ in the line element (cf. Eq. (1)) instead of the advanced time $v$ \cite{papadopoulos, nerozzi,barreto}. 

Whenever mass is absorbed by the original black hole, a new black hole is formed with an increased mass. In our formulation, this process leads to the creation of an apparent horizon characterized by \( \beta \rightarrow \infty \), which interrupts the numerical integration. During the evolution of the gravitational waves, we observed a decrease in Bondi mass, consistent with the Bondi formula.

%The central issue we addressed is the relationship between the extracted or absorbed mass and the initial amplitude \( \epsilon \) that characterizes the initial amount of mass-energy surrounding the black hole.

The central issue we have addressed is to establish a relationship between the extracted or absorbed mass and the initial amplitude \( \epsilon \), which characterizes the initial mass-energy distribution around the black hole. We found that the scaling law (37), with an exponent $\nu \approx 2$, accurately fits the numerical data generated from various initial data families. Furthermore, we show that a generalization of this power law, represented by the nonextensive distribution (38), provides a slightly better fit to the numerical data.

Finally, we initiated an investigation into the effects of placing a spherical reflecting barrier around the black hole at $r=R_0$, with the condition $R_0 > 2M_0$. While this setup is somewhat artificial, it introduces a novel boundary condition for the collapse of gravitational waves. As illustrated in Fig. 3, weak gravitational waves are entirely radiated to infinity due to the reflecting barrier, leaving the original black hole with a Bondi mass equal to $M_0$. However, above a critical strength, the collapse of gravitational waves leads to the formation of a horizon, where part of the gravitational waves is radiated away, and the remaining portion, along with the reflector, falls into the black hole, resulting in a new black hole with increased mass. Future work will extend this investigation to critical collapse scenarios beyond spherical symmetry and with mass gap \cite{crespo_deoliveira}.

\acknowledgments

We thank Conselho Nacional de Desenvolvimento Cient\'ifico e Tecnol\'ogico (CNPq) and Funda\c c\~ao Carlos Chagas Filho de Amparo \`a Pesquisa do Estado do Rio de Janeiro (FAPERJ)
(Grant No. E-26/200.774/2023 Bolsas de Bancada de Projetos (BBP)).

\appendix

\section{The field equations}

The field equatios in dimensionless variables are listed below:

\begin{widetext}
	\begin{eqnarray}
		\beta_{,\eta} &=& \displaystyle{\frac{1}{2} (\eta+1)\gamma_{,\eta}^2}, \label{eq4}\\
		\nonumber \\
		Q_{,\eta} &=& \displaystyle{2(\eta+1)^2\,\bigg\{(\eta+1)^2\bigg[\frac{\beta}{(\eta+1)^2}\bigg]_{,\eta\theta} - \frac{\left(\sin^2 \theta\,\gamma\right)_{,\eta \theta}}{\sin^2 \theta} + 2 \gamma_{,\eta} \gamma_{,\theta}\bigg\}}, \label{eq5}\\
		\nonumber \\
		U_{,\eta}&=&\frac{Q}{(\eta+1)^4}\mathrm{e}^{2(\beta-\gamma)} \\
		\nonumber \\ 
		V_{,\eta} &=& -\frac{1}{4}(\eta+1)^4 \mathrm{e}^{2(\gamma-\beta)}\left(U\right)_{,\eta}^2 + \frac{\big[(\eta+1)^4 \sin \theta\,U \big]_{,\eta\theta}}{2 (\eta+1)^2 \sin \theta} + \mathrm{e}^{2(\beta-\gamma)}\bigg[1-\frac{(\sin \theta\,\beta_{,\theta})_{,\theta}}{\sin \theta} +\gamma_{,\theta\theta} + 3 \cot \theta\,\gamma_{,\theta} \nonumber \\
		\nonumber \\
		&&- (\beta_{,\theta})^2 - 2\gamma_{,\theta}(\gamma_{,\theta}-\beta_{,\theta})\bigg], \label{eq6} \\
		\nonumber \\
		\Gamma_{,\tilde{u}\eta} &=&\frac{1}{4(\eta+1)} \Bigg\{2(\eta+1)\gamma_{,\eta}V - (\eta+1)^2\left[ 2\gamma_{,\theta}U + \sin \theta\,\left(\frac{U}{\sin \theta}\right)_{,\theta}\right]\Bigg\}_{,\eta} - (\eta+1)\frac{(\gamma_{,\eta}U\sin \theta)_{,\theta}}{2\sin \theta} \\
		\nonumber \\
		&&+ \frac{1}{8}(\eta+1)^3\mathrm{e}^{2(\gamma-\beta)} (U_{,\eta})^2 
		+ \frac{\mathrm{e}^{2(\beta-\gamma)}}{2(\eta+1)}\bigg[(\beta_{,\theta})^2 + \sin \theta \left(\frac{\beta_{,\theta}}{\sin \theta}\right)_{,\theta}\bigg],   
	\end{eqnarray}
\end{widetext}

\noindent where $\gamma=\Gamma/(1+\eta)$ and we removed the bars for convenience.

\section{Spectral approximation and the basis functions}

We presented the spectral approximation of the function $\beta$ in Eq. (25). The corresponding spectral approximations for the remaining metric functions are 

\begin{eqnarray}
	&&U(\tilde{u},\eta,x) =
	\sum_{k}^{N_\eta^{(\beta)}}\,\sum_{j=0}^{N_x^{(\beta)}}\,\hat{U}_{kj}(\tilde{u}) \Psi_k^{(\beta)}(\eta) P_{2j+1}(x), \\
	\nonumber \\
	&&Q(\tilde{u},\eta,x) =
	\sum_{k}^{N_\eta^{(\beta)}}\,\sum_{j=0}^{N_x^{(\beta)}}\,\hat{Q}_{kj}(\tilde{u}) \eta TL_k(\eta) P_{2j+1}(x), \\
	\nonumber \\
	&&V(\tilde{u},\eta,x) = \eta +
	\sum_{k}^{N_\eta^{(S)}}\,\sum_{j=0}^{N_x^{(S)}}\,\hat{V}_{kj}(\tilde{u}) \Psi_k^{(\Gamma)}(\eta) P_{2j}(x), \\
	\nonumber \\
	&&\Gamma(\tilde{u},\eta,x)=
	\sum_{k}^{N_\eta}\,\sum_{j=0}^{N_x}\,\hat{\Gamma}_{kj}(\tilde{u}) \Psi_k^{(\Gamma)}(\eta) P_{2j}(x).
\end{eqnarray}

\noindent The radial basis functions $\Psi_k^{(\beta)}(\eta)$ and $\Psi_k^{(\Gamma)}(\eta)$ are defined by:

\begin{eqnarray}
	\Psi_k^{(\Gamma)}(\eta) &&= \frac{1}{2}\left(TL_{k+1}(r)+TL_{k}(\eta)\right) \nonumber \\
	\\
	\Psi_k^{(\beta)}(\eta) &&= a_1TL_{k+2}(\eta)+a_2TL_{k+1}(\eta)+a_3TL_k(\eta),  \nonumber 
\end{eqnarray} 

\noindent where  $a_1=-(2k+1)/4(2k+3)$, $a_2=(k+1)/(2k+3)$, and $a_3=-1/2$.

\end{document}